\documentstyle[epsf,aps,multicol]{revtex}

\begin{document}
\draft

\title{\Large\bf Nature of the vortex-glass order in
strongly type-II superconductors}

\author{Hikaru Kawamura}

\address{Department of Earth and Space Science,
Faculty of Science, Osaka University, Toyonaka, 560-0043,
Japan}

\maketitle

\begin{abstract}
The stability and the critical properties
of the three-dimensional vortex-glass order in
random type-II superconductors with point disorder
is investigated in the unscreened limit
based on a lattice {\it XY\/} model
with a uniform field. 
By performing equilibrium Monte Carlo simulations for
the system with periodic boundary conditions,
the existence of a stable
vortex-glass order
is established in the unscreened limit.
Estimated critical exponents are compared
with those of the gauge-glass model.
\end{abstract}


\begin{multicols}{2} \narrowtext

In spite of extensive studies for a decade, the question
of nature of the thermodynamic phase diagram of random high-$T_c$
superconductors has remained unsettled. In zero field, the possibility
of an exotic thermodynamic phase with broken time-reversal symmetry,
called the chiral-glass phase, has been discussed\cite{KawaLi1}. Even more
attention has been paid to the in-field properties.
For superconductors with point disorder,
possible existence of a thermodynamic phase
called the vortex-glass (VG) phase, where the vortex
is pinned on long length scale by randomly distributed
point-pinning centers,
was proposed \cite{Fisher}. In such a VG state,
the phase of the condensate wavefunction is frozen in time
but  randomly in space, with a
vanishing linear resistivity $\rho _L$.
It is a truly superconducting state separated from
the vortex-liquid phase with a nonzero $\rho _L$ via a continuous
VG transition.

Since cuprate high-$T_c$ superconductors are extremely type-II
superconductors where the London penetration depth $\lambda $ is
much longer than the coherence length, it is important
to clarify first whether the proposed VG state really exists in the
type-II, unscreened limit $\lambda \rightarrow \infty $. Indeed,
stability of the hypothetical VG state
has been studied quite extensively by numerical model simulations
\cite{Huse,Reger,WY,OY,HK,Wallin,OT}.
Many have been based on a highly simplified model called the
the gauge-glass model.
Previous simulations on the three-dimensional (3D) gauge-glass
model gave mutually consistent results that a continuous VG transition
occurred at a finite temperature characterized by the critical
exponents, $\nu \simeq 1.3$, $\eta \simeq -0.5$, $z\simeq 4-5$
\cite{Huse,Reger,WY,OY}.

The gauge-glass model has some drawbacks \cite{Huse}.
It is a spatially isotropic model
without a net field threading the system, in contrast to the reality.
Furthermore,  source of
quenched randomness is artificial.
The gauge-glass model is a random flux model where the quenched
randomness occurs in the phase factor associated with the flux.
In  reality, the quenched component of the flux
is uniform, nothing but the external field, and the quenched
randomness occurs in the superconducting coupling or the pinning
energy.
It remains unclear whether these simplifications underlying
the gauge-glass model really unaffect the basic physics of the
VG ordering in 3D.

Recently, several simulations
were performed beyond the gauge-glass model\cite{HK,Wallin,OT}.
The present author studied the type of the
lattice {\it XY\/}
model where the above limitations of the gauge-glass model were
cured\cite{HK}.
While the VG state was found to be stable, the estimated critical
exponents, particularly $\nu\simeq 2.2 $,
differed significantly from those of the gauge-glass model,
posing a possibility that the gauge-glass model lied in a different
universality class.  However, due to the
effect of employed free boundary conditions,
the estimated critical exponents might possibly be subject to large
surface effect.  Vestergren
{\it et al\/} studied a random pinning model which took care of
the above limitations of the gauge-glass model in a different way, to obtain
a finite-temperature VG  transition characterized by the
exponents, $\nu\simeq 0.7,\ z\simeq 1.5$\cite{Wallin},
which differed significantly from either those of Ref.\cite{HK}
or from those of the gauge-glass model\cite{Reger,WY,OY}.
Olsson and Teitel claimed
on the basis of their simulations on the lattice {\it XY\/}
model with weak disorder that the VG order was not stable
even in the unscreened limit\cite{OT}. Thus, once one tries to go beyond
the gauge-glass model, the present theoretical situation  seems
quite confused.

In the present paper, I study the lattice {\it XY\/} model
of Ref.\cite{HK}, but now with applying {\it periodic boundary conditions\/},
to overcome the finite-size effect originated from surface. 
The  Hamiltonian considered is
\begin{equation}
H = - \sum _{<ij>} J_{ij}\cos (\theta _i-\theta _j-A_{ij}),
\end{equation}
where $\theta _i$ is the phase of the condensate
at the $i$-th site of a simple cubic lattice with
$N=L^3$ sites,
and the sum
is taken over all nearest-neighbor pairs.
$A_{ij}$ is a link variable associated with the vector potential
due to uniform external magnetic field of intensity $h$ applied in
the $z$-direction. In the Landau gauge, it is given by
${\bf A}_i$=($A_i^x$,$A_i^y$,$A_i^z$)=(0,$hi_x$,0), where
$1\leq i_x\leq L$ denotes the $x$-coordinate of the site $i$.
Quenched randomness occurs in the superconducting coupling
$J_{ij}$ which is assumed to
be an independent
random variable uniformly distributed between [0,$2J$],
$J>0$ being a typical coupling strength.  
I impose periodic boundary conditions
in all directions in order to eliminate surface `spins' which
might contaminate the bulk critical behavior. 
The field intensity is chosen to be $h=2\pi/4$ ($f=1/4$).
The lattice sizes are taken to be multiples of four,
{\it i.e.\/}, $L=8,12,16$ and 20.

Simulation is performed based on the exchange MC method,
where the systems at neighboring temperatures are occasionally
exchanged\cite{HN}. Equilibration is checked by monitoring the
stability of the results
against at least three-times longer runs for a subset of samples.
Sample average is taken over
980 ($L=8$ and 12), 248 ($L=16$) and 200 ($L=20$) 
independent bond realizations.

We run two independent sequences of  systems
(replica 1 and 2) in parallel, and
compute a complex overlap $q$ between the local superconducting
order parameters of the two replicas $\psi_i^{(1,2)}\equiv
\exp (i\theta _i^{(1,2)})$,
\begin{equation}
  q = \frac{1}{N}\sum_{i}\psi_{i}^{(1)*}\psi_{i}^{(2)},
\end{equation}
where the summation is taken over all $N=L^3$ sites.
In terms of the overlap $q$, the VG order parameter and the
Binder ratio
is calculated by
\begin{equation}
 q^{(2)}=[<|q|^2>],\ \ \ \ 
 g=2-\frac{[\langle |q|^4\rangle]}
    {[\langle  |q|^2\rangle]^2},
\end{equation}
where $\langle\cdots\rangle$ represents the thermal average
and $[\cdots ]$
represents the average over bond disorder.

By inspecting the
spatial pattern of the vortex snapshot and by calculating
the Fourier amplitude of the spatial distribution of vortices, it is
checked that
no periodic vortex-lattice order is formed in the system.

The size and temperature dependence
of the calculated Binder ratio
is shown in Fig.1(a).
As can be seen from the figure, except for the smallest size $L=8$,
$g$ for different $L\geq 12$ cross 
at $T/J=0.82\pm 0.02$, indicating that the VG transition occurs
at a finite temperature. The data
show a rather clear splay-out, in contrast to the near marginal
merging behavior observed for the case of free boundary conditions\cite{HK}.
In contrast to the case of the chiral-glass order in the
{\it XY\/} spin glass\cite{KawaLi2}, 
$g$ does not exhibit a negative dip characteristic
of one-step-like replica symmetry breaking.

Further evidence of a finite-temperature transition is obtained from the
mean-square currents\cite{Reger}. 
The current
in the $\mu$ direction $I_\mu$  ($\mu=x,y,z$) is defined by
\begin{equation}
I_\mu = \frac{1}{L}
\sum _{<ij>\parallel \mu}J_{ij}\sin (\theta_i-\theta_j-A_{ij}),
\end{equation}
where the sum is taken over all
nearest-neighbor bonds along the $\mu$ direction. 
We calculate both the longitudinal
(along the applied field) and the transverse (perpendicular to the applied
field) mean-square currents given by 
\begin{equation}
I_L^2=[<I_z>^2],\ \ I_T^2=[<I_x>^2+<I_y>^2].
\end{equation}
\begin{figure}[h]
\epsfxsize=\columnwidth\epsfbox{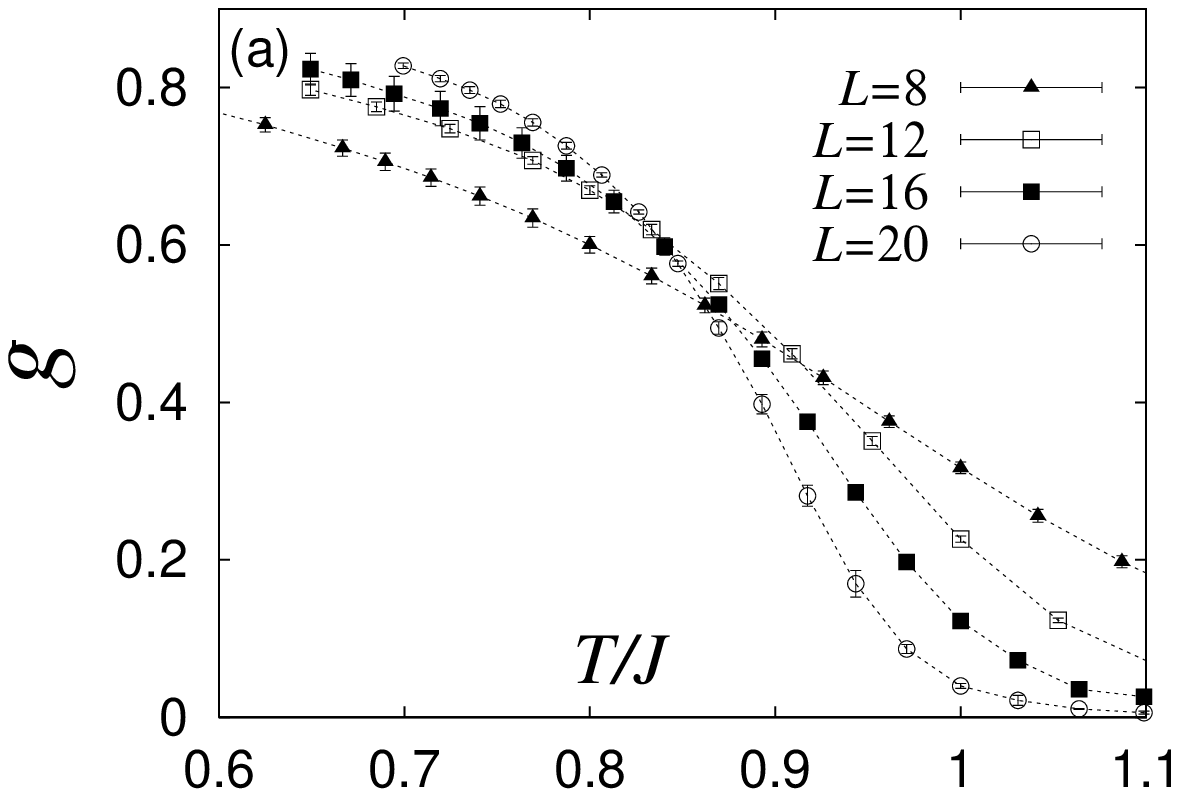}
\epsfxsize=\columnwidth\epsfbox{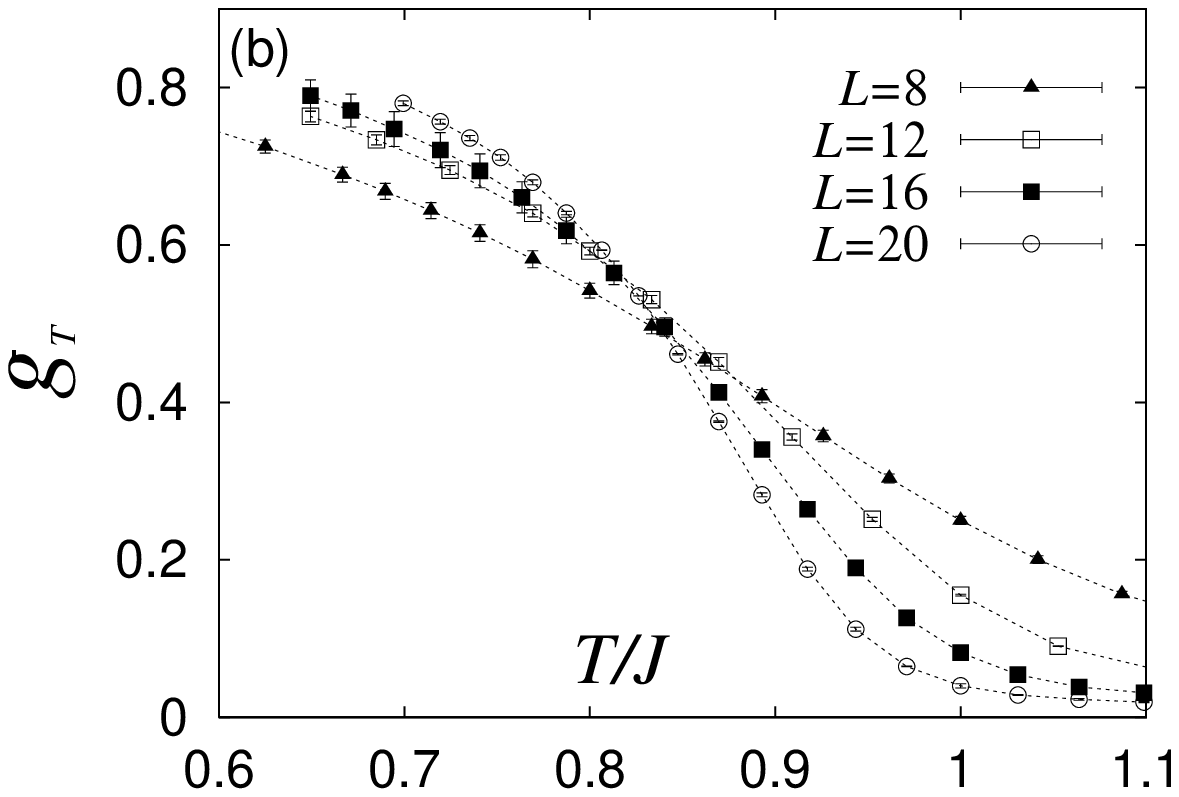}
\caption{
Temperature and size dependence of
(a) the bulk Binder ratio,
and of (b)  the transverse Binder ratio.
}
\end{figure}

As can be seen from Fig.2(a), 
the transverse mean-square current for $L\geq 12$ exhibits a crossing at
$T/J= 0.81\pm 0.02$, 
close to the crossing temperature of the Binder ratio.
The longitudinal mean-square current also exhibits a crossing behavior,
as is shown in Fig.2(b). 
Apparently, finite-size effects are severer in
$I_L^2$  than in $I_T^2$, 
as the crossing for smaller sizes occurs at relative high
temperature, which shifts down rapidly to lower
temperature
with increasing $L$, {\it i.e.\/}, $T_{{\rm cross}}/J>1.1$ 
($L=8$ and 12), $\simeq 1.00$
($L=12$ and 16), and $\simeq 0.89$ ($L=16$ and 20). 
Aside from this strong finite-size correction, 
the observed behavior
seems consistent with the occurrence of a single bulk transition
at $T/J\simeq 0.81$ as suggested from the
transverse mean-square current. 

In order to further confirm that  both the transverse and
longitudinal spatial components order
simultaneously,
I also compute the transverse VG order parameter and the
transverse Binder ratio\cite{HK}. These quantities are 
defined in terms of
the layer-overlap $q'_k$  for the $k$-th $xy$-layer of
the lattice,
\begin{equation}
  q'_k = \frac{1}{L^2}\sum_{i\in k}
\psi_{i}^{(1)*}\psi_{i}^{(2)},
\end{equation}
where $i\in k$ means the sum over all sites belonging to the
$k$-th layer, by
\begin{equation}
  q^{(2)}_T=\frac{1}{L}\sum_k [<|q'_k|^2>],\ \ 
  g_T=2-\frac{\sum_k [\langle |q'_k|^4\rangle]}
    {\sum_k [\langle  |q'_k|^2\rangle]^2}.
\end{equation}
\begin{figure}[h]
\epsfxsize=\columnwidth\epsfbox{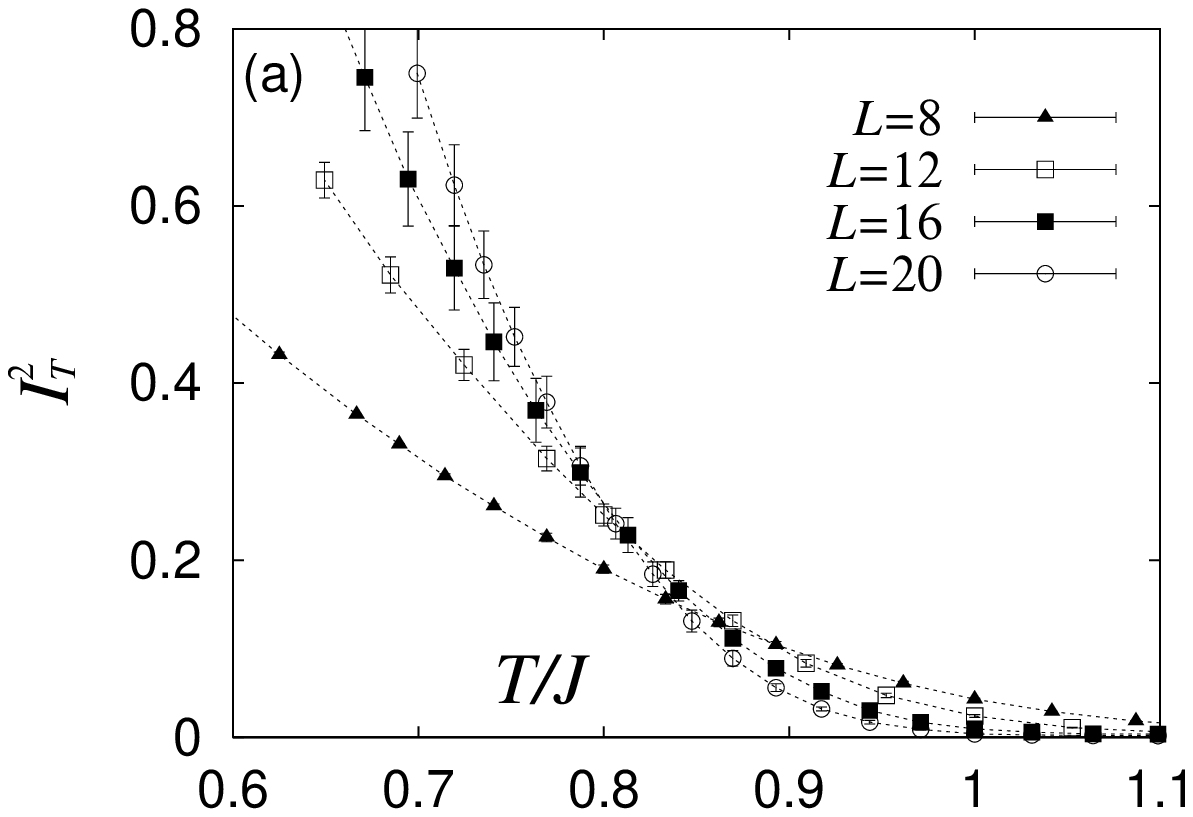}
\epsfxsize=\columnwidth\epsfbox{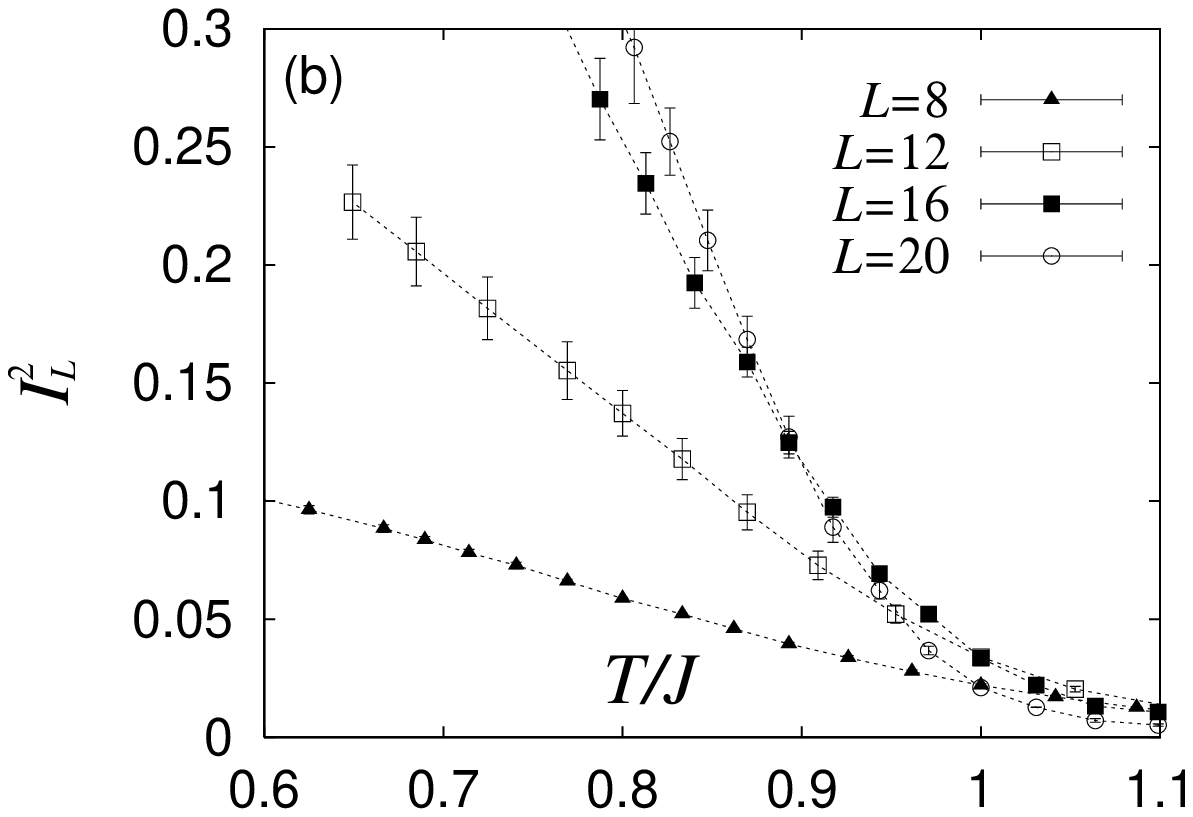}
\caption{
Temperature and size dependence of (a)
the transverse and (b) the longitudinal mean-square currents.
}
\end{figure}
The calculated $g_T$ is shown in Fig.1(b).
As can be seen from the figure, $g_T$
shows a behavior quite similar to $g$,
exhibiting 
a clear crossing at $T/J=0.81\pm 0.02$.
This also indicates that the model exhibits a single bulk
VG transition where both the transverse and longitudinal components
order simultaneously.
 
Next, we turn to the critical properties of the model.
Standard finite-size scaling analysis performed for the bulk Binder
ratio, with setting $T_g/J=0.82$,  yields 
$\nu=1.2\pm 0.3$, as shown in the inset of  Fig.3 where
$t\equiv (T-T_g)/T_g$. Then, 
from the scaling of the order parameter $q^{(2)}$,
the critical-point-decay exponent is estimated to be $\eta=-0.5\pm 0.1$:
see Fig.3.
The estimate of $\nu$ is also corroborated by a finite-size scaling analysis
of the transverse mean-square current $I_T^2$,  
which, with $T_g/J=0.81$, yields $\nu=1.0\pm 0.2$. The 
corresponding finite-size scaling plot is given in Fig.4. 
(Unfortunately,
meaningful scaling analysis is not feasible for $I_L^2$  
due to the rapid shift of the crossing points with $L$.)
In order to examine the possibility of anisotropic scaling,
finite-size scaling of $g_T$ and
$q^{(2)}_T$ (see the inset of Fig.4) is also performed. I get
$\nu =1.2\pm 0.3$ and $\eta =-0.5\pm 0.1$,
which agree  within the errors
with $\nu $ and $\eta $ determined from $g$ and $q^{(2)}$.
No sign of anisotropic scaling is thus found.
Combining all these estimates,
I finally quote $\nu =1.1\pm 0.2$, $\eta =-0.5\pm 0.1$.
\begin{figure}[h]
\epsfxsize=\columnwidth\epsfbox{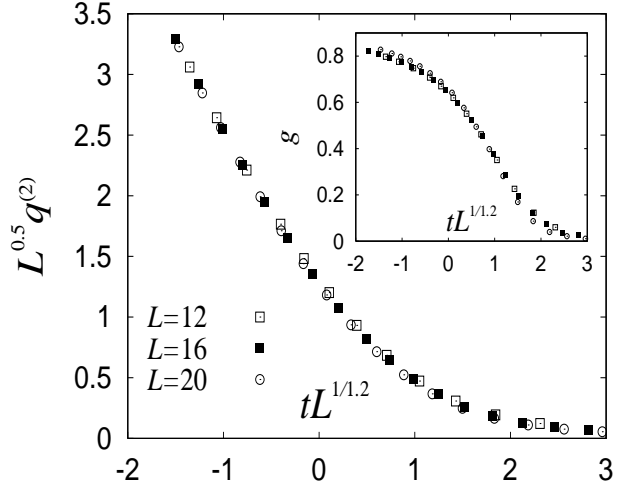}
\caption{
Finite-size scaling plot of the VG order parameter (main panel)
and of the bulk Binder ratio (inset), with $T_g/J=0.82$,
$\nu=1.2$ and $1+\eta=0.5$.
}
\end{figure}
\begin{figure}[h]
\epsfxsize=\columnwidth\epsfbox{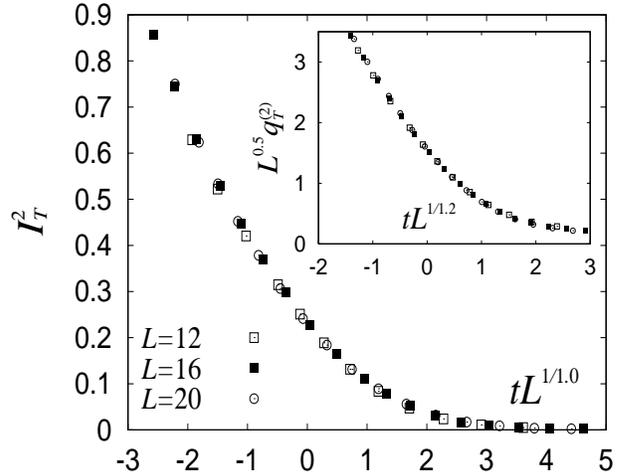}
\caption{
Finite-size scaling plot of the transverse mean-square current (main panel)
and of the transverse VG order parameter (inset), with $T_g/J=0.81$.
Best values of exponents are taken to be $\nu=1.0$ for $I_T^2$,
while $\nu=1.2$ and $1+\eta=0.5$ for $q^{(2)}_T$.
}
\end{figure}
On comparing the exponent values obtained here with those obtained for
the system with free boundary conditions\cite{HK}, one sees that they
differ considerably. In the range of lattice sizes studied here,
the application of either periodic or free boundary
significantly influences the estimates of critical
exponents. One naturaly expects that 
periodic boundary conditions give better
estimates for the bulk exponents, since no surface 'spin' exists there
which might contaminate the bulk critical properties.

If one compares the present estimates of exponents with
those of the the gauge-glass model (with periodic boundary), the
values of both $\nu$  and $\eta $ are 
compatible with each other, {\it i.e.,} $\nu=1.1\pm 0.2$ and 
$\eta=-0.5\pm 0.1$ 
for the present model vs. $\nu=1.39\pm 0.20$ and $\eta=-0.47\pm 0.07$ 
for the gauge-glass model\cite{OY}. This coincidence might
suggest that the present model
belongs to the same universality class as the gauge-glass model.
It thus appears that 
the differences in the form of quenched randomness and in the spatial
anisotropy due to external fields are irrelevant to the critical
properties of the VG transition.

Finally, I wish to refer to the possible effect of screening 
in real VG ordering. 
In real type-II superconductors including high-$T_c$ superconductors, 
the penetration
depth is large, but of cource is not infinite. As several calculations
on the gauge-glass model and other random {\it XY\/} model
have constantly suggested,  
finite screening effect might
destabilize the stable VG state\cite{HK,Young,Rieger}. 
Since high-$T_c$
superconductors are strongly type-II superconductors, 
this screening-induced rounding of a sharp VG transition
is a weak effect, visible
only at very close to $T_g$ of the unscreened system. 
With use of the present estimate of $\nu$,
a rough estimate of such a rounding (or crossover) temperature $T_\times$ 
may be obtained. 
Screening effect would be visible when the coherence
length $\xi \approx \xi_0t^{-\nu}$ ($\xi_0$ is the zero-temperature
coherence length and $t \equiv (T-T_g)/T_g$) 
grows comparable to the  zero-temperature
penetration depth $\lambda_0$. Since the ratio $\lambda_0/\xi_0$
is of order $10^2$ in high-$T_c$ superconductors, the crossover temperature
is of order $t_\times \sim 10^{-2}$. It means that 
one has to approach $T_g$ as close as $t\sim 10^{-2}$
in order to see the screening-induced rounding.
In other words, in the temperature range outside $t_\times$,
the critical behavior of the unscreened 
system is expected to be observed experimentally
\cite{Kwok}.  

In summary, the VG ordering of strongly type-II superconductors
with point disorder
is investigated
based on a lattice {\it XY\/} model with a uniform field.
The occurrence of  a finite-temperature
VG transition
is established in the unscreened limit. The estimated critical exponents
$\nu$ and $\eta$ are close to the corresponding gauge-glass values,
suggesting that the present model belong to the same universality
class as the gauge-glass model.

The author is thankful to Dr.S. Teitel for useful discussion. He is
also indebted to Dr.J. Lidmar for his comment on the 
estimate of critical exponents.
The numerical calculation was performed on the Hitachi SR8000
at the supercomputer center, ISSP, University of Tokyo.

\medskip
\noindent
{\it Note added}

In the course of preparing this manuscript, the author learned
that Olsson also made a MC simulation of the 3D VG order in
the unscreened limit based on a
lattice {\it XY\/} model, 
with a diffierent choice of the random-coupling distribution
[cond-mat/0301624]. 
Finite-temperature VG transition
with an exponent $\nu=1.50\pm0.12$ was observed,
roughly being consistent with our present result.
Critical exponents similar to our present values
were aslo reported very recently by Lidmar based on a different type of
VG model [cond-mat/0302577].

\end{multicols}

\end{document}